\documentclass[useAMS,usenatbib,usegraphicx,fleqn]{mn2e}
\usepackage{times}
\usepackage{balance}
\usepackage{amsmath}
\usepackage{amssymb}

\newcommand{\stars}{\textsc{stars}}
\newcommand{\Msun}{\hbox{$\rm\thinspace \text{M}_{\sun}$}}
\newcommand{\Rsun}{\hbox{$\rm\thinspace \text{R}_{\sun}$}}
\newcommand{\Lsun}{\hbox{$\rm\thinspace L_{\sun}$}}

\newcommand{\yr}{\rm\thinspace yr}
\newcommand{\Myr}{\rm\thinspace Myr}

\newcommand{\Msunpyr}{\hbox{$\Msun\yr^{-1}\,$}}
\newcommand{\K}{\rm\thinspace K}
\newcommand{\cm}{\rm\thinspace cm}
\newcommand{\cmsq}{\hbox{$\cm^2\,$}}
\newcommand{\pcmcu}{\hbox{$\cm^{-3}\,$}}
\newcommand{\g}{\rm\thinspace g}
\newcommand{\pg}{\hbox{$\g^{-1}\,$}}
\newcommand{\gpcm}{\hbox{$\g\cm^{-3}\,$}}

\newcommand{\Htu}{\text{H}_2}
\newcommand{\rb}{r_\text{B}}
\newcommand{\rs}{r_\text{S}}
\newcommand{\Lbh}{L_\text{BH}}
\newcommand{\Mbh}{M_\text{BH}}
\newcommand{\dMbh}{\dot{M}_\text{BH}}
\newcommand{\sci}[2]{#1\times10^{#2}}
\newcommand{\bracfrac}[2]{\left(\frac{#1}{#2}\right)}

\newcommand{\tdif}[2]{{\frac{d #1}{d #2}}}

\newcommand{\shorteq}[1]{
\begin{equation}
#1
\end{equation}}
\newcommand{\mchead}[1]{\multicolumn{2}{c}{#1}}

\newcommand{\aap}{A\&A}
\newcommand{\aapr}{A\&AR}
\newcommand{\aj}{AJ}
\newcommand{\apj}{ApJ}
\newcommand{\apjl}{ApJL}
\newcommand{\apjs}{ApJS}
\newcommand{\araa}{ARA\&A}
\newcommand{\mnras}{MNRAS}
\newcommand{\nat}{Nat}
\newcommand{\pasj}{PASJ}

\pagerange{2751--2762}
\volume{414}
\pubyear{2011}
\date{Accepted 2011 February 23. Received 2011 January 27; in original form 2010 August 25}

\begin{document}

\title{The structure and evolution of quasi-stars}
\author[W. H. Ball et al.]
{Warrick H. Ball,\thanks{E-mail: wball@ast.cam.ac.uk}
Christopher A. Tout, Anna N. \.Zytkow,
and John J. Eldridge\\ 
Institute of Astronomy, The Observatories, Madingley Road, Cambridge CB3 0HA}
\maketitle

\begin{abstract}
  The existence of bright quasars at high redshifts implies that
  supermassive black holes were able to form in the early
  Universe. Though a number of mechanisms to achieve this have been
  proposed, none yet stands out. A recent suggestion is the formation
  of quasi-stars, initially stellar-mass black holes accreting from
  hydrostatic giant-like envelopes of gas, formed from the monolithic
  collapse of pre-galactic gas clouds. In this work, we modify the
  Cambridge \stars{} stellar evolution package to construct detailed
  models of the evolution of these objects. We find that, in all of
  our models, the black hole inside the envelope is able to reach
  slightly more than one-tenth of the total mass of the system before
  hydrostatic equilibrium breaks down. This breakdown occurs after a
  few million years of evolution. We show that the mechanism which
  causes the hydrostatic evolution to end is present in polytropic
  models. We also show that the solutions are highly sensitive to the
  size of the inner boundary radius and that no physical solutions
  exist if the inner boundary is chosen to be less than about $0.3$ of
  the Bondi radius.
\end{abstract}

\begin{keywords}
accretion, accretion discs -- black hole physics -- quasars: general
\end{keywords}

\section{Introduction}
\label{sint}

Over the last decade, high-redshift surveys have revealed the
existence of bright quasars at redshifts $z\ga 6$
\citep{fan+06,jiang+08,willott+10}. Such observations imply the
existence of supermassive black holes (SMBHs) with masses
$\Mbh\ga10^9\Msun$ less than $10^9\yr$ after the Big Bang. It remains
unclear how these objects became so massive so quickly. Substantial
and ongoing research has led to various possibilities, each subject to
a number of uncertainties \citep[see the recent review
by][]{volonteri10}.

One popular possibility is that the SMBHs that power high-redshift
quasars grew from smaller \emph{seed} black holes (BHs) that were the
remnants of the first generation of stars. The Big Bang did not
produce significant amounts of elements heavier than helium
so these first stars are expected to have been
composed purely of hydrogen and helium. They have a number of features
which distinguish them as a population and are thus termed
\mbox{\emph{population III}} (pop-III) stars \citep[see][for a
review]{bromm+04}.

Pop-III stars are thought to have formed at the centres of smaller
dark matter (DM) haloes, with virial temperatures in the range
$10^3\la T_\text{vir}/\K\la10^4$, in which the gas cooled through
molecular hydrogen emission. These stars had masses in the range
$100\la M/\Msun\la1000$. Models predict that those with masses of more
than about $260\Msun$ undergo pair-instability supernovae and leave BH
remnants of about half of their progenitor masses
\citep*{fryer+01}. As the DM haloes continue to merge and grow, so the
seed BH accretes gas from its surroundings, settles to the centre of
the halo and grows in mass.

How long would it take such a seed BH to grow to $10^9\Msun$? If the
BH were able to accrete constantly at its Eddington-limited rate, then
a $100\Msun$ BH would take about ${\sci{7}{8}\yr}$ to grow into a
$10^9\Msun$ SMBH for a typical radiative efficiency of the accretion
($\epsilon=0.1$). This is barely the age of the Universe at $z=6$ and
already brings the scenario into doubt. In addition, the assumptions
of sufficient fuel and constant Eddington accretion are
weak. \citet{johnson+07} find that the supernova marking the formation
of a BH causes a delay of order $100\Myr$ before the BH begins to
accrete efficiently. \citet*{milosavljevic+09} find that the accretion
on to a BH from a dense protogalactic cloud is self-limiting and the
maximum accretion rate is about 32 per cent of the Eddington-limited
rate. \citet{king+06} have shown that, if accretion occurs in small,
frequent, randomly-aligned episodes, the time constraint is relaxed
because the BH spin stays low on average, reducing the radiative
efficiency. In the case of monolithic collapse from a protogalactic
cloud, it is not clear that the material reaches the centre with
random alignments, although this may well be the case for subsequent
merger-driven accretion.

It thus appears that pop-III seeds cannot grow sufficiently rapidly to
become the BHs that power high-redshift quasars. However, if the
first generation of stars pollute the interstellar medium
with a sufficient metal content, then the second generation of stars
resembles modern populations. During hierarchical mergers, gas builds
up in the cores of haloes and undergoes fragmentation, forming dense
stellar clusters \citep*{clark+08}. The dense environment of these
stars can lead to frequent stellar collisions and thence either directly
to a massive BH or to a massive star which leaves a massive BH as its
remnant \citep{devecchi+09}.

\citet*{spolyar+08} suggested that DM annihilation inside pop-III 
stars can significantly heat them. This could delay the star's
arrival on the main sequence and allow it to grow to a larger mass
\citep{freese+08}. Once the DM is exhausted, normal evolution
would proceed but larger seed BH masses are possible than for typical
pop-III stars.

A separate branch of possible paths to early SMBH formation stems from
the direct collapse of supermassive clouds ($M>10^5\Msun$) at the
centres of DM haloes with $T_\text{vir}>10^4\K$. If $\Htu$ formation
is suppressed, the only coolant for primordial gas is atomic
hydrogen. The gas is only able to cool efficiently to about $10^4\K$
\citep{tegmark+97} and cosmological simulations show
that it condenses into thick, pressure-supported discs that are
gravitationally stable \citep*{regan+09a,wise+08a}. The cores of such
discs are then able to collapse into isolated structures with masses
that exceed $10^4\Msun$.

The formation of these massive objects hinges on the suppression of
$\Htu$ formation, at least until particle densities around
$10^5\pcmcu$ are reached, whereafter collisional dissociation is
sufficient to prevent fragmentation \citep*{schleicher+10}. A popular
mechanism for this suppression is photodissociation by an ionizing UV
background. \citet*{shang+10} estimate that the necessary specific
intensity exceeds what is expected on average in the relevant
epoch. However, \citet{dijkstra+08} suggest that the inhomogeneous
distribution of ionizing sources suppresses $\Htu$ formation in a
fraction of haloes. \citet{spaans+06} argue that the self trapping of
Ly$\alpha$ radiation during the collapse keeps the temperature above
$10^4\K$. The importance and effect of $\Htu$ remains uncertain but
the work cited above indicates a number of paths which overcome this
hurdle.

The further central collapse of the discs described above requires
transport of angular momentum. \citet{lodato+06} claim that, for
haloes with $T_\text{vir}/T_\text{gas}<1.8$, angular momentum is
transported by local gravitational instabilities such that the disc
maintains a state of marginal stability. \citet*{begelman+06} instead
argue that, when a critical threshold of rotational support is
achieved, the clouds become unstable to the formation of bars, which
transport angular momentum outwards and material inwards on dynamical
timescales. If the gas remains able to cool, then the process can be
repeated and a cascade of bars is formed. This is the
\emph{bars-within-bars} mechanism described by \citet*{shlosman+89}
and seen in the simulations of
\citet{wise+08a}. \citet{begelman+09} go further to show how it
also suppresses fragmentation by maintaining supersonic
turbulence and thus relaxes the requirement of the absence of $\Htu$.

Ultimately, if a large body of gas is able to condense in the centre
of the DM halo, it can take one of two forms. On the one hand it may
form a very massive star. If the star is sufficiently massive, it is
prone to general-relativistic instabilities and collapses into a BH
with\,90 per cent of the progenitor mass \citep{shapiro04}. If the
total mass of the star exceeds about $10^5\Msun$ the collapse occurs
before the end of the main sequence. If the total mass is smaller but
still exceeds $\sci{3.4}{4}\Msun$, then it collapses during helium
burning \citep*{fricke+71,bond+84}. If the total mass is smaller
still then the star becomes pair-unstable after core helium burning
and leaves a BH of about half the progenitor mass \citep{ohkubo+06}.

If the rate of mass infall is much higher then the envelope of the
star does not reach thermal equilibrium during the lifetime of the
star \citep{begelman10}. In this case, the core collapses after
hydrogen burning is complete. The structure that remains after core
collapse is a stellar-mass BH embedded within a giant-like envelope,
or a \emph{quasi-star} \citep{begelman+06}. The attractive feature of
a quasi-star is that the accretion rate on to the BH is limited by the
Eddington rate of the entire object, which is initially much larger than 
that of the BH alone. The excess energy is carried away by convection.

Quasi-stars are the subject of this paper. \citet*[][hereinafter BRA08]
{begelman+08} examined their structure using analytic estimates
and basic numerical results. We use a fully fledged
stellar evolution code to study quasi-stars' structure and evolution
without many of the assumptions made by BRA08.

In \mbox{Section \ref{smeth}}, we describe the Cambridge \stars{} code and
the modifications that we have made. In \mbox{Section \ref{sfid}},
we present the results of a fiducial model and make a short
theoretical diversion to explain some of the model's behaviour in
\mbox{Section \ref{slimit}}. We then present results with varied parameters
in \mbox{Section \ref{smore}} and compare our models with some of the
estimates made by BRA08 in \mbox{Section \ref{sbra}}. We explore the 
sensitivity of our models to the inner boundary radius in 
\mbox{Section \ref{sbound}} before concluding in \mbox{Section \ref{sconc}}.

\section{Method}
\label{smeth}

The models that we report on in this paper were computed with a
modified version of the Cambridge \stars{} code. The original
code was written by \citet{eggleton71,eggleton72,eggleton73}
and it has been substantially updated by \citet{pols+95} and
\citet{eldridge+04}. The principal modification to the code is 
the change to the interior boundary conditions. Normally,
stellar evolution codes solve for the interior boundary conditions
$r,\,m,\,L_r=0$, where 
$r$ is the radial co-ordinate,
$m$ is the mass within a radius $r$ and
$L_r$ is the luminosity through the sphere of radius $r$.
We replace these conditions with a prescription for the
BH's interaction with the surrounding gas as described below.

To derive suitable conditions we presume that the pressure in the
envelope is dominated by the radiation from the accreting
BH. \citet{loeb+94} showed that a radiation-dominated fluid in
hydrostatic equilibrium, which is not generating energy, must become
convective. Thus the envelope of our quasi-star is approximated by a
gas with polytropic index $n=3$. We derive the boundary conditions
below under this presumption but in the calculations we use the
adiabatic index ($\gamma=(d\log p/d\log\rho)_S$) determined
self-consistently by the equation-of-state module in the model.

\subsection{Radial boundary condition}

The radius of the inner boundary of the envelope must be the point at
which some presumption of the code breaks down. In particular, we
choose a radius within which the gas is no longer expected to be in
hydrostatic equilibrium. A reasonable choice is the \emph{Bondi radius}
$\rb$, the radius at which the thermal energy of the fluid particles
equals their gravitational potential energy.
\shorteq{\frac{1}{2}mc_s^2=\frac{Gm\Mbh}{\rb}\label{bondidef1}\text{,}}
so
\shorteq{\rb=\frac{2G\Mbh}{c_s^2}\label{bondidef2}\text{,}}
where 
$m$ is the mass of a test particle, 
$c_s=\sqrt{\gamma p/\rho}$ is the adiabatic sound speed,
$\Mbh$ is the mass of the BH
and $G$ is the gravitational constant.
Inside this boundary we expect the BH's gravity to overcome the
thermal motion of the fluid. The average radial velocity of the 
fluid is presumed to be inward, although convective motions
may result in local motions both towards and away from the BH.

\subsection{Mass boundary condition}

For the mass boundary condition, consider the mass of the gas inside
the cavity defined by the Bondi radius, \mbox{equation (\ref{bondidef2})}. 
By definition
\shorteq{M_\text{cav}=\int^{\rb}_{\rs}4\pi r^2\rho(r)dr\label{mcav}\text{,}}
where $\rs$ is the Schwarzschild radius. Using a general
relativistic form of the equation introduces terms of order $\rb/\rs$
\citep{thorne+77} which we ignore because all our models have
$\rs\ll\rb$. 

To determine the mass of gas inside the cavity, we must make some
assumption about the density profile of the material therein because
the code does not model this region. The quasi-star is supported by 
radiation pressure and is expected to radiate near its Eddington 
limit. The Eddington limit for the entire quasi-star is much greater 
than the same limit for the BH alone and excess flux drives bulk 
convective motions. The radial density 
profile of the accretion flow then depends on whether angular 
momentum is transported outward or inward. In the former case, 
the radial density profile is proportional to $r^{-\frac{3}{2}}$ 
whereas, in the latter case, it is proportional to $r^{-\frac{1}{2}}$
\citep{narayan+00}. We presume that the viscosity due to small scale 
magnetic fields is sufficiently large to transport angular momentum
outwards even if convection transports it inwards and thus take
$\rho(r)\propto r^{-\frac{3}{2}}$.

Given the density $\rho(\rb)=\rho_0$ at the inner boundary, the 
density profile must be
\shorteq{\rho(r)=\rho_0\bracfrac{r}{\rb}^{-\frac{3}{2}}\text{.}}
We evaluate \mbox{equation (\ref{mcav})}, presuming $\rs\ll\rb$, to find
\shorteq{M_\text{cav}=\frac{8\pi}{3}\rho_0\rb^3\label{mcavint}\text{.}}
In a radiation-dominated $n=3$ polytrope, 
the pressure and density are related by
\shorteq{P=\bracfrac{k}{\mu m_\text{H}}^\frac{4}{3}
\bracfrac{3(1-\beta)}{a\beta^4}^\frac{1}{3}\rho^\frac{4}{3}
= K\rho^\frac{4}{3}}
\citep{eddington18}, where 
$k$ is Boltzmann's constant, 
$\mu$ is the mean molecular weight of the gas, 
$m_\text{H}$ is the mass of a hydrogen atom and 
$\beta=P_\text{g}/P$ is the ratio of gas pressure to total pressure.
Taking the adiabatic sound speed to be $c_s=\sqrt{4P/3\rho}$,
evaluating the Bondi radius using \mbox{equation (\ref{bondidef2})} and
substituting into \mbox{equation (\ref{mcavint})}, we obtain
\shorteq{M_\text{cav}=\frac{8\pi}{3}\bracfrac{3G\Mbh}{2K}^3\text{,}}
which we can use to consider the importance of the mass
inside the cavity. \citet{fowler64} gives
$\beta=4.3(M_*/\Msun)^{-\frac{1}{2}}/\mu$, where $M_*$ is the total mass
of the object. For a totally ionized
mixture of 70 per cent hydrogen and 30 per cent helium,
$\mu=0.615$. For a total quasi-star mass of $10^4\Msun$, as in our
fiducial results, we find that $M_\text{cav}=\Mbh$ when
$\Mbh\approx390\Msun$. So we must include $M_\text{cav}$ in the mass
boundary condition.

Thus, we set the boundary condition for the mass co-ordinate to
\shorteq{M_0=\Mbh+M_\text{cav}\label{mbc}\text{,}}
where $M_\text{cav}$ is given by equation (\ref{mcavint}). 
In general, different accretion modes give different
prescriptions for the density inside the cavity and thus
different cavity masses. In our fiducial results, we have assumed that
the density inside the cavity is as described above. In Section 
\ref{sscavprop} we construct a model presuming that 
$\rho(r)\propto r^{-\frac{1}{2}}$ and find that changing the density 
profile inside the cavity has little effect.

\subsection{Luminosity boundary condition}

The final boundary condition that must be included is the
luminosity. The luminosity is determined by the mass accretion rate
through the relationship
\shorteq{\Lbh=\epsilon\dot{M}c^2\label{lum1}\text{,}} where $c$ is the
speed of light, $\dot{M}$ is the rate of mass flow across the base of
the envelope and $\epsilon$ is the radiative efficiency, the fraction
of accreted rest mass that is released as energy. This fraction is
lost from the system as radiation so the total mass of the quasi-star
decreases over time. The actual rate of accretion on to the BH is
$\dMbh\equiv(1-\epsilon)\dot{M}$, the amount of infalling matter less
the radiated energy. The luminosity condition is related to the actual
BH accretion by
\shorteq{\Lbh=\frac{\epsilon}{1-\epsilon}\dMbh c^2=\epsilon'\dMbh c^2\text{.}}
We thus implicitly assume that the material travels from the base of the 
envelope to the event horizon within one timestep. The material actually 
falls inward on a dynamical timescale so this condition is already implied 
by the presumption of hydrostatic equilibrium.

To specify the mass accretion rate from the inner boundary we begin
with the adiabatic Bondi accretion rate \citep{bondi52},
\shorteq{\dot{M}_\text{Bon}=4\pi\lambda_c\frac{(G\Mbh)^2}{c_s^3}\rho_0
\label{dmbhbondi}\text{,}}
where $\lambda_c$ is a factor that depends on the adiabatic
index $\gamma$ as described by equation (18) of \citet{bondi52}.
For the case of $\gamma=4/3$, $\lambda_c=1/\sqrt{2}$.
Almost all of this flux is carried away from the
BH by convection. BRA08 point out that the maximum convective flux in
the material is $pc_s$ so that the maximum luminosity is

\begin{align}
L_\text{con,max}&=4\pi \rb^2 pc_s \\
&=\frac{4}{\gamma}\pi\rb^2c_s^3\rho \\
&=\frac{1}{\gamma\lambda_c}\dot{M}_\text{Bon}c_s^2\text{.}
\end{align}
In order to limit the luminosity to the convective maximum,
the accretion rate is reduced by a factor
$c_s^2/\gamma\lambda_c\epsilon' c^2$. We presume that the actual
convective flux is some fraction of the maximum computed above and
thus implement the mass accretion rate
\shorteq{\dMbh=4\pi\frac{\eta}{\epsilon'\gamma}\frac{(G\Mbh)^2}{c_sc^2}\rho\text{,}
\label{dmbhbc}}
where $\eta$ represents the convective efficiency. 
For the fiducial run, we take $\eta=\epsilon=0.1$.

\subsection{Further assumptions}

We conclude this section with a brief discussion of the remaining
presumptions in the code. The temperatures and densities at the base
of the envelope are generally too low for meaningful energy generation
from nuclear reactions to take place, so we did not solve the chemical
evolution equations in the results presented here. To confirm that our
approximation is sound, preliminary runs were performed with nuclear
reactions included. No discernible difference was found in the
results.

This does not immediately exclude the possibility of nuclear reactions
occurring inside the hydrodynamic region. Using a temperature profile
$T\propto r^{-1}$ \citep{narayan+00}, we estimated the composition
changes due to the pp-chains presuming complete mixing down to
$10^{-4}\,\rb$. We found no significant change to the H and He
abundances and conclude that the associated energy generation is also
negligible. Although the temperatures in these regions are well over
$10^8\K$, the densities in the region inside $\rb$ are typically only
a few $\gpcm$.

We have not performed detailed calculations that consider the
CNO-cycle. Prior to its collapse to a BH, the stellar core is
expected to synthesize sufficient CNO to maintain hydrostatic
equilibrium through the hydrogen-burning phase
\citep{begelman10}. Most of these metals fall on to the black hole
unless convection is established with the envelope quickly at core
collapse. If the core abundance is mixed, it is largely diluted by the
pristine material in the envelope.

Our model of the interior also neglects any loss of heat via neutrino
emission. To establish whether this is reasonable, we estimated the
total neutrino loss rate using the analytic estimates of
\citet{itoh+96} and integrated them over the interior region for the
models in the fiducial run. If the flow extends all the way to the
innermost stable circular orbit (ISCO) of a non-rotating BH,
$r_\text{ISCO}=6\rs$, we find the neutrino losses are at most 6 per
cent of the total luminosity. If the flow is truncated at
$2r_\text{ISCO}$, the neutrino losses come to less than 0.02 per
cent. Such losses would effectively decrease the radiative efficiency
but the accretion rate is principally determined by the convective
efficiency. We thus believe our model envelopes remain stable against
catastrophic neutrino losses.

If the advection of material across the event horizon is faster than
the time taken by radiation to diffuse out, then energy can be lost to
the BH. Such radiation trapping reduces the support for the infalling
material and enhances the accretion rate. \citet{begelman78}
determined an approximate expression for this increase which is
supported by the numerical results of \citet{flammang84}. We
recomputed the fiducial run with the appropriate factor included and
found no difference in the results.

For now, we have not included non-spherical effects of rotation inside
the hydrodynamic region. Such effects may be important.
Convection-dominated accretion flows are expected to have moderate
angular dependence but do not support mass ejection. If convection
maintains constant specific angular momentum then even Keplerian
rotation at a few hundred $\rs$ becomes dynamically insignificant at
$\rb$. We expect any outflows to be impeded by the material at or near
the base of the envelope and that convective turbulence preserves
approximate spherical symmetry in the vicinity of the inner radius. We
intend to introduce an approximate treatment of rotation to the
envelope models in future work.

The \stars{} code includes a detailed equation-of-state package
\citep{eggleton+73,pols+95} which computes the ionization states of H
and He and the contribution of $\Htu$. The gas is everywhere
approximated well by a sum of radiation pressure and the ideal gas law
but the ionization state of the material has a substantial effect on
the structure of the envelope.

\section{Fiducial results}
\label{sfid}

We begin the exposition of our results by selecting a run which we
shall use to demonstrate the qualitative features of a quasi-star
envelope's structure and evolution. We subsequently experiment with
various properties of the model to indicate how the behaviour is
affected by such changes.

The results presented in this section describe a model quasi-star 
with initial total mass (BH, cavity gas and envelope) $M_*=10^4\Msun$.
The BH initially has mass $0.0005M_*=5\Msun$ but the evolution does 
not depend on this fraction. The gas is uniformly composed of 0.7 
hydrogen and 0.3 helium by mass. The envelope is allowed to relax 
to thermal equilibrium before the BH begins accreting.

\begin{table*}\begin{minipage}{160mm}
\caption{Properties of the fiducial model for increasing values of $\Mbh$.
The first and last entries correspond to the initial and final models in the run, 
respectively. Density profiles are plotted in Fig. \ref{ltrop+dinv}.}
\label{fidprop}
\begin{tabular}{@{}crcr@{}lccr@{}lr@{}lr@{}lr@{}l@{}}
\hline
$t$        &$\Mbh$ &$\dMbh$           &\mchead{$M_\text{cav}$}
& $L_*$         &$\rho_0$&\mchead{$T_0$}&\mchead{$T_\text{eff}$}& 
\mchead{$\rb$}  & \mchead{$R_*$} \\
/$10^6\yr$&/\Msun&/$10^{-4}\Msunpyr$&\mchead{/\Msun}&/$10^8\Lsun$
&/\gpcm &\mchead{/$10^5\K$}&\mchead{/$10^3\K$}& 
\mchead{/$100\Rsun$}&\mchead{/$10^4\Rsun$} \\
\hline
0.00  &5      &2.13   &    0&.00 &3.48&$\sci{8.71}{-5}$ &40&.8&14&.3&0&.0166&0&.303\\
0.51  &100    &1.79   &    3&.83 &2.92&$\sci{5.47}{-8}$ &3&.54&5&.22&3&.66&2&.09 \\
1.03  &200    &2.08   &   24&.9  &3.40&$\sci{1.30}{-8}$ &2&.23&4&.77&11&.1&2&.70 \\
2.23  &500    &2.96   &  241&    &4.83&$\sci{2.42}{-9}$ &1&.33&4&.55&41&.2&3&.54 \\
3.70  &1000   &3.70   & 1359&    &6.05&$\sci{6.49}{-10}$&0&.88&4&.49&113& &4&.07 \\
4.23  &1194   &3.53   & 3360&    &5.81&$\sci{3.71}{-10}$&0&.71&4&.51&185& &3&.96 \\
\hline
\end{tabular}
\end{minipage}\end{table*}

\subsection{Structure}
\label{ssstruc}

\begin{figure}
\includegraphics[width=84mm]{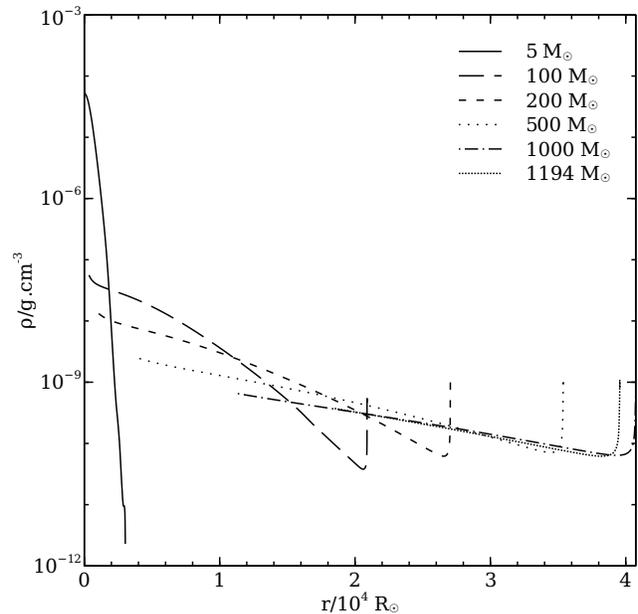}
\caption{Plot of density against radius for models in the fiducial run with 
$\Mbh/\Msun=5,100,200,500,1000$ and $1194$ (see Table \ref{fidprop}). 
At the base of the envelope the density profile steepens because
of the steeper pressure gradient required to balance the BH gravity. 
In the outer layers the density is inverted, as discussed by BRA08.
Note that the initial model has inner radius of $1.66\Rsun$ which is
too small to be seen.}
\label{ltrop+dinv}
\end{figure}

In the model, the luminosity is approximately equal to the Eddington
luminosity at the boundary of the innermost convective layer. The
accretion rate varies between about $\sci{1.8}{-4}$ and
$\sci{3.7}{-4}\Msunpyr$ as the convective boundary moves. The details
of the variation are described in Section \ref{ssevol}. A corollary of
the self-limiting behaviour is that the only major effect of changing
the material composition is to change its opacity. This in turn
changes the Eddington limit and therefore the accretion rate but the
envelope structure remains almost entirely unchanged. In the
convective regions, the envelope has an adiabatic index of about
$1.34$ (corresponding to a polytropic index $n\approx2.90$) confirming
the expectation that the envelope is approximated by an $n=3$
polytrope. The boundaries of the convective regions depend on the
ionization state of the gas but, for most of the evolution, all but
the outermost few $10\Msun$ are convective.

Fig. \ref{ltrop+dinv} shows a sequence of density profiles of the
envelope when $\Mbh/\Msun=5$, $100$, $200$, $500$, $1000$ and $1194$ (further
parameters are listed in Table \ref{fidprop}). These profiles
demonstrate two features of the envelope structure. The first is the
central condensation which can be seen from the slight rise in the
density at the innermost radii. This feature is clearer at smaller BH
masses. It is caused by the lack of pressure support at the inner
boundary. In order to remain in hydrostatic
equilibrium, the equations require
\shorteq{\left.\tdif{P}{r}\right|_{\rb}=-\frac{G\rho M_0}{\rb^2}}
at the boundary, so the pressure gradient steepens
and the density gradient follows. \citet{huntley+75} called such 
structures \emph{loaded polytropes}.

The second feature, apparent in all but the first density profile in
Fig. \ref{ltrop+dinv}, is the density inversion in the outer
layers. It appears once the photospheric temperature $T_\text{surf}$
drops below about $8000\K$. From then, the surface opacity increases
owing to hydrogen recombination. The Eddington luminosity falls and
the star's luminosity apparently exceeds the Eddington limit. It is
well known that hydrostatic models can sustain this super-Eddington
luminosity through a density inversion \citep{langer97}. As an
additional check, we calculated the volume-weighted average of
$3\gamma-4$ and found it to be positive, indicating dynamical (but not
pulsational) stability \citep[][Section 27.3b]{cox+68}.

\subsection{Evolution}
\label{ssevol}

The sequence of density profiles in Fig. \ref{ltrop+dinv} shows the
interior density decreasing over time. If we consider equation
(\ref{dmbhbc}), ignoring constants and using $P\propto\rho^\gamma$, then
\shorteq{\dMbh\propto\Mbh^2\rho^\frac{3-\gamma}{2}\text{.}} 
The accretion rate $\dMbh$ is held approximately constant by the
Eddington limit and $\Mbh$ is always increasing. Thus, for any
reasonable adiabatic index ($\gamma<3$), $\rho$ decreases at the inner
boundary. Initially, the density decreases relatively rapidly and the
envelope expands. The expansion occurs owing to the opacity peak at
the surface due to hydrogen. The rate of change of the surface radius
is at most about $0.1\Rsun\yr^{-1}$ which is five orders of magnitude
smaller than the free-fall velocity. These models are thus still in
hydrostatic equilibrium.

\begin{figure}
\includegraphics[width=84mm]{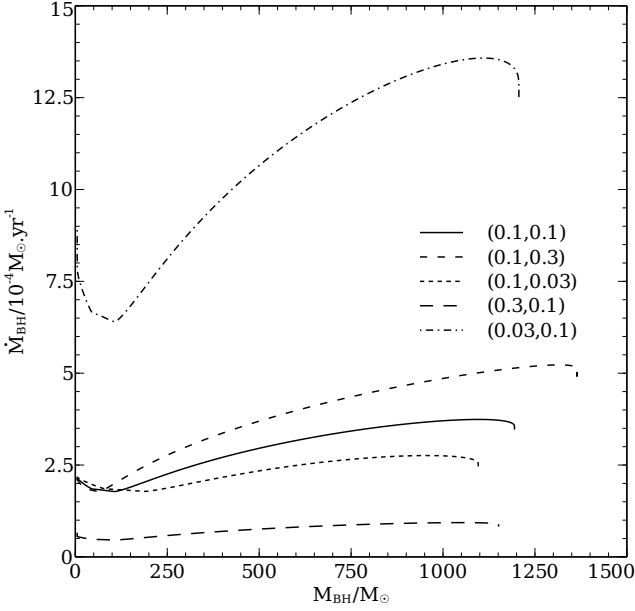}
\caption{Plot of BH accretion rate $\dMbh$ against BH mass $\Mbh$ for
various pairs of values of the radiative and convective efficiencies
($\epsilon,\eta$). The fiducial values ($0.1,0.1$) correspond to the
solid line. As expected, changing the radiative efficiency changes
$\dMbh$ but leaves the overall structure unaffected. Decreasing $\eta$ 
causes the envelope to be hotter and denser in order to achieve the 
same luminosity shifting the discontinuities in the gradient of the
accretion rate to later times.}
\label{epseta}
\end{figure}

\begin{figure}
\includegraphics[width=84mm]{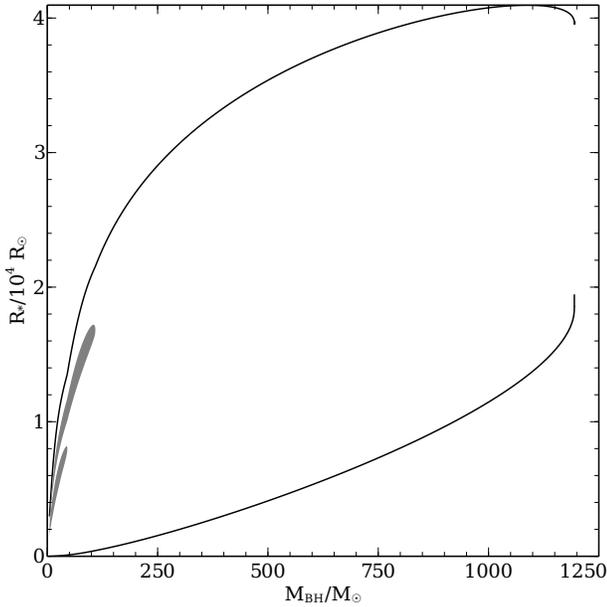}
\caption{Radial location of convective boundaries and the Bondi radius
for the fiducial run. The shaded regions are the radiative parts of
the envelope. The outer solid lines show the extent of the
hydrostatic envelope. The qualitative change in the locations of
convective boundaries causes the discontinuities in the gradient of the 
accretion rate that are seen in Fig. \ref{epseta}. The outermost layer 
of the envelope is radiative (as it should be) but it is too narrow to 
be seen here.}
\label{convboun}
\end{figure}

Fig. \ref{epseta} shows the accretion rate on to the BH as a function
of its mass. Fig. \ref{convboun} shows the locations of convective 
boundaries as a function of BH mass and demonstrates
how the rapid changes of the BH accretion rate while $\Mbh<120\Msun$
coincide with the disappearance of radiative regions. 
The disappearance is due to
the decreasing density throughout the envelope.

Before the end of the evolution the accretion rate achieves a local
maximum. At the same time, the photospheric temperature reaches a
local minimum and the envelope radius a maximum (see
Fig. \ref{convboun}). We do not have a simple explanation for this but
some insight is offered in \mbox{Section \ref{ssuv}}. Evolution
beyond the results shown here is impossible. The code reduces the
timestep below the dynamical timescale indicating that it cannot
construct further models that satisfy the structure equations.

\subsection{Termination of the run and subsequent evolution}

For the fiducial run, the evolution terminates when
$\Mbh=1194\Msun$. The physical reason for this upper limit remains
elusive but we have made some progress in understanding it using a
modified version of the Lane-Emden equation (see \mbox{Section
 \ref{slimit}}). The existence of the limit is certainly robust as it
is does not depend on the total mass of the quasi-star over at least
two orders of magnitude (see \mbox{Section \ref{ssinitmass}}) nor on
whether the envelope mass changes in time (see \mbox{Section
 \ref{ssml}}). At the end of the run the cavity contains a further
$3360\Msun$. Under our assumptions, some of this material is already
moving towards the BH and may become part of it. If the BH accretes
all the mass in the cavity, its final mass would be
$\Mbh\approx4554\Msun$, nearly half of the total mass of the original
quasi-star. Presuming the BH accretes at its Eddington-limited rate,
this growth would take about $57\Myr$.

What actually happens to the material in the cavity after the end of
the hydrostatic evolution? Because we do not model it, we can only
speculate. It appears that the entire envelope might be swallowed but
it is not certain that this should be the case. The accretion flow is
convective so there must be a combination of inward and outward
flowing material within the Bondi radius. In the theoretical limit for
a purely convective flow, the accretion rate is zero
\citep{quataert+00} and about half of the material is moving inward
and half outward. If the flow is sustained then we might expect at
least half of the cavity mass to be accreted. On the other hand the
flow structure might change completely. The infalling material could
settle into a disc and drive disc winds or jets so that the overall
gain in mass is relatively small. The envelope is also convective and
we expect equal masses of gas to be moving inward and
outward. Hydrostatic equilibrium is presumably failing so the dynamics
might change drastically here, too. Recently, \citet{johnson+10}
modeled the accretion on to massive BHs formed through direct
collapse. They assumed that the BH accretes from a multi-colour black
body disc after its quasi-star phase and found that, once the BH mass
exceeds about $10^4\Msun$, the accretion rate decreases due to
radiative feedback. This result supports the case for a substantial
decrease in the BH growth rate if a thin disc forms after the
quasi-star phase but additional growth can occur during the transition
to a new structure.

\section{Final BH mass limit}
\label{slimit}

We have not yet found a simple explanation for the existence 
of the upper limit to the BH mass in a quasi-star. In this section, we
explore some theory that confirms the existence of such a
limit and points towards possible mechanisms.

\subsection{Loaded polytropes}
\label{sslp}

Consider the equations of hydrostatic equilibrium and mass
conservation truncated at some radius $r_0$ and loaded with some mass
$M_0$ interior to that point. The equations are
\shorteq{\tdif{P}{r}=-\frac{Gm\rho(r)}{r^2}}
and
\shorteq{\tdif{m}{r}=4\pi r^2\rho(r)}
with the central boundary conditions $\left.m\right|_{r_0}=M_0$,
where $r_0$ and $M_0$ are fixed. We scale the pressure and density
using the usual polytropic assumptions
\shorteq{P=K\rho^{1+\frac{1}{n}}}
and
\shorteq{\rho=\rho_0\theta^n\text{.}}
We define the dimensionless radius by
\shorteq{r=\alpha\xi\text{,}}
where
\shorteq{\alpha^2=\frac{(n+1)K}{4\pi G}\rho_0^{\frac{1}{n}-1}\text{.}}
We scale the mass interior to a sphere of radius $r$ by defining 
\shorteq{\phi(\xi)=\frac{m}{4\pi\rho_0\alpha^3}}
\citep{huntley+75}. The non-dimensional form of the equations is then
\shorteq{\tdif{\theta}{\xi}=-\frac{1}{\xi^2}\phi\label{dtheta}}
and
\shorteq{\tdif{\phi}{\xi}=\xi^2\theta^n\label{dphi}}
with boundary conditions $\theta(\xi_0)=1$ and
$\phi(\xi_0)=\phi_0$ (where, by definition, $\xi_0=r_0/\alpha$).\footnote
{If one takes $\xi_0=\phi_0=0$, differentiates equation 
(\ref{dtheta}) and substitutes for $d\phi/d\xi$ using equation 
(\ref{dphi}), one arrives at the usual Lane-Emden equation.}

We can express $\phi_\text{BH}$ (the scaled BH mass) in terms of the
inner radius by rescaling equation (\ref{bondidef2}) as follows.
\begin{align}
\xi_0=\frac{r_0}{\alpha}
&=\frac{2G}{\alpha}\frac{\Mbh}{c_s^2} \\
&=\frac{2G}{\alpha}4\pi\rho_0\alpha^3\phi_\text{BH}\frac{n}{(n+1)K\rho_0^\frac{1}{n}}\\
&=2n\alpha^2\phi_\text{BH}\frac{4\pi G}{(n+1)K\rho_0^{\frac{1}{n}-1}} \\
&=2n\phi_\text{BH}\text{.}
\end{align}
Similarly, we can derive the following relation for
$M_\text{cav}$ from equation (\ref{mcavint}).
\begin{align}
\phi_\text{cav}
&=\frac{\frac{8\pi}{3}\rho_0(\alpha\xi_0)^3}{4\pi\rho_0\alpha^3}\\
&=\frac{2}{3}\xi_0^3\text{.}
\end{align}
The scaled mass and radius boundary conditions are now related by
\shorteq{\phi_0\equiv\phi(\xi_0)=\frac{1}{2n}\xi_0+\frac{2}{3}\xi_0^3\text{.}}
Thus, for a given polytropic index $n$, we can choose a
value $\xi_0$ and integrate the equations. 

\begin{figure}
\includegraphics[width=84mm]{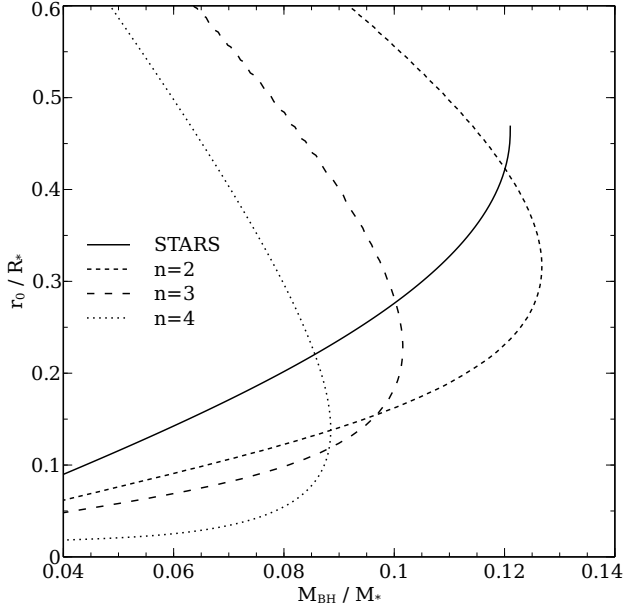}
\caption{Plot of the ratio of BH mass against total mass against inner 
envelope radius to outer envelope radius. The double-valued curves 
correspond to polytropic solutions (with $n=2,3$ and $4$). The monotonic 
curve (solid line) represents the fiducial results. 
The polytropic models show the existence of an upper limit to the BH mass 
ratio. The fiducial results reach a similar limit, but the mass ratio cannot 
decrease, so the evolution terminates. }
\label{matlab}
\end{figure}

We integrated a sequence of solutions for $n=3$ and found that 
there exists a maximum value of $\phi_\text{BH}/\phi_*=0.102$,
where $\phi_*$ is the scaled total mass of the quasi-star.
The maximum occurs when $\xi_0=0.995$. 
We investigated polytropic indices between $2$ and $4$ and found a similar
limit in all cases. For $n=2$ we found a maximum $\phi_\text{BH}/\phi_*=0.127$ when
$\xi_0=1.012$ and for $n=4$ the maximum was $\phi_\text{BH}/\phi_*=0.089$
at $\xi_0=0.968$. Fig. \ref{matlab} shows plots of the curves of the
ratio of inner to outer envelope radius ($\xi_0/\xi_*$ in scaled variables,
where $\xi_*$ is the outer radius of the envelope) against the BH
fractional mass ($\phi_\text{BH}/\phi_*$ in scaled variables)
for $n=2$, $3$ and $4$ together with our results for the fiducial model. 
The maximum mass ratio is clear in each curve. In principle, further 
hydrostatic solutions probably exist along the sequence computed by 
\stars{} but they require that the BH mass decreases. 

The conclusion we draw is that the models produced by the code come to
a halt because no further realistic hydrostatic solutions can be found along
the sequence. The discussion in this section, even though it does not immediately 
offer an explanation of why, thus indicates that such a limit
does truly exist. The limit is robust in the sense that it exists at
approximately the same ratio for all masses and does not depend on
details such as opacity or energy transport (as long as the envelope is
convective). The main dependence in real models is due to variation in
the polytropic index.

\subsection{The $U$--$V$ plane}
\label{ssuv}

It is known that \citep[see][Section 4.8]{chandra39}, if one has determined 
a solution $\theta(\xi)$ to the Lane-Emden equation, then $\theta'(\xi')
\equiv C^\frac{2}{n-1}\theta(C\xi)$ is also a solution for an
arbitrary constant $C$. By selecting appropriate variables that are
invariant to such a transformation, one can find all the related
solutions in a single calculation. Though a number of appropriate
variables exist, many authors \citep[such as][]{kimura81,horedt87} 
use $U$ and $V$, defined below. Converting to the new variables 
also allows the analysis of solutions of the polytropic equation 
that do not extend to $\xi=0$ or contain the entire mass of the 
envelope. Physical solutions are restricted to positive $U$ and 
$V$. The first quadrant of the plane that they form is a useful 
tool for analyzing the behaviour of solutions.

We begin by defining
\shorteq{U=\frac{\xi^3\theta^n}{\phi}}
and
\shorteq{V=\frac{\phi}{\xi\theta}\text{.}}
Let us differentiate the logarithms of these variables with
respect to $\xi$.
\begin{align}
\frac{1}{U}\tdif{U}{\xi}
&=\frac{3}{\xi}+\frac{n}{\theta}\tdif{\theta}{\xi}-\frac{1}{\phi}\tdif{\phi}{\xi} \\
&=\frac{3}{\xi}-\frac{n\phi}{\theta\xi^2}-\frac{\xi^2\theta^n}{\phi} \\
&=\frac{1}{\xi}(3-nV-U)
\end{align}
and
\begin{align}
\frac{1}{V}\tdif{V}{\xi}
&=-\frac{1}{\xi}-\frac{1}{\theta}\tdif{\theta}{\xi}+\frac{1}{\phi}\tdif{\phi}{\xi} \\
&=-\frac{1}{\xi}+\frac{\phi}{\theta\xi^2}+\frac{\xi^2\theta^n}{\phi} \\
&=\frac{1}{\xi}(-1+U+V)\text{.}
\end{align}
Dividing these two equations eliminates $\xi$ and allows us
to write the first-order equation
\shorteq{\tdif{V}{U}=-\frac{V}{U}\left(\frac{U+V-1}{U+nV-3}\right)\text{.}}
Polytropes that have zero mass and non-zero density
at $\xi=0$ begin at $(U,V)=(3,0)$.

\begin{figure}
\includegraphics[width=84mm]{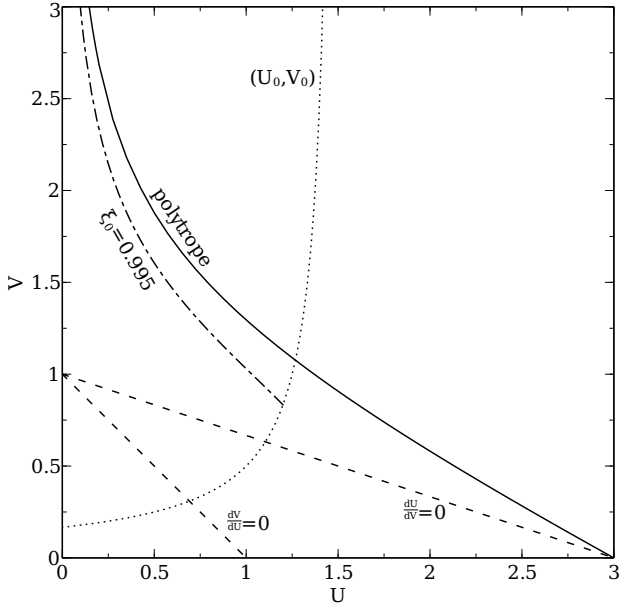}
\caption{Some features of the $U$--$V$ plane for $n=3$. The dashed
lines correspond to limiting values of $dV/dU$, as indicated on the
plot. The dotted line shows the curve of central boundary conditions
for the loaded polytropes described in \mbox{Section \ref{sslp}}, with BH mass
increasing towards the top and right. The solid line shows the standard
$n=3$ polytrope. The dash-dotted line is the loaded polytrope
corresponding to the largest BH-envelope mass ratio. The intersection
of the $dU/dV=0$ line and the curve of initial values corresponds well
with the maximum accretion rate in in Fig. \ref{epseta}. It is
interesting that the final model takes initial values approximately
halfway between this point and the intersection with the $n=3$
polytrope.}
\label{uv}
\end{figure}

Using the scaled boundary conditions in the previous section, we can
determine the curve in the $U$--$V$ plane which corresponds to the
interior values of our models. The relevant initial points are given by
\shorteq{U_0=\frac{\xi_0^3\theta_0^n}{\phi_0}=\xi_0^2
\left(\frac{1}{2n}+\frac{2}{3}\xi_0^2\right)^{-1}}
and
\shorteq{V_0=\frac{\phi_0}{\xi_0\theta_0}=\frac{1}{2n}+\frac{2}{3}\xi_0^2\text{.}}
We can eliminate $\xi_0$ to give the curve of initial points
in the $U$--$V$ plane.
\shorteq{V_0=\frac{3}{2n}\left(3-2U_0\right)^{-1}\text{.}}
Different BH masses correspond to different points along
this curve. A solution for a particular BH mass is then a 
curve going from the relevant point on the $(U_0,V_0)$ curve
towards $U=0$.

Fig. \ref{uv} shows a number of features in the $U$--$V$ plane for
$n=3$. Along the $dV/dU=0$ or $dU/dV=0$ lines, the solutions are
horizontal or vertical, respectively, in the $U$--$V$ plane. All
solutions bounded below by the $dU/dV=0$ have decreasing $U$ and
increasing $V$ everywhere.

How do the points of interest correspond to our models? The
intersection of the $dU/dV=0$ line and the $(U_0,V_0)$ curve
corresponds to $\xi_0=0.835$, for which the BH-envelope mass ratio is
$0.0953$. This appears to correspond to the point at which the
accretion rate achieves its final local maximum. The dash-dotted 
line indicates the solution to the equations in
\mbox{Section \ref{sslp}} that has the largest BH-envelope mass ratio. 
The curve's initial point appears to be about halfway between the
intersections of the $(U_0,V_0)$ curve with the $dU/dV=0$ and with the
polytropic solution. Though the mechanism behind the mass limit is
not yet known, the analysis in this section shows that the limit is
reproduced in the polytropic approximation, so the relevant physics 
is contained within mass conservation and hydrostatic equilibrium.

\section{Further results}
\label{smore}

In \mbox{Section \ref{sfid}}, we established the basic qualitative structure
and evolution of the quasi-star envelope. In this section, we explore
their dependence on some of the parameters of the model. 

\subsection{Radiative and convective efficiencies}

We first vary the accretion rate by adjusting the parameters
$\epsilon$ and $\eta$. Fig. \ref{epseta} shows the accretion history
against BH mass for a number of parameter choices. Because the
luminosity always settles on the same convection-limited rate, changing
$\epsilon$ only rescales the accretion rate through equation
(\ref{lum1}) and has no effect on the structure. The final BH mass
and intermediate properties are the same. The only difference is
that the evolution takes longer for larger values of $\epsilon$.

Increasing the convective efficiency $\eta$ allows a greater flux to be
radiated. Because it is a limiting factor, a larger value of
$\eta$ allows a larger accretion rate for given interior
conditions. In order to establish the same overall luminosity, the
envelope must be less dense. This explains the different times at which
the discontinuities in the gradient of the accretion rate appear.

\begin{figure}
\includegraphics[width=84mm]{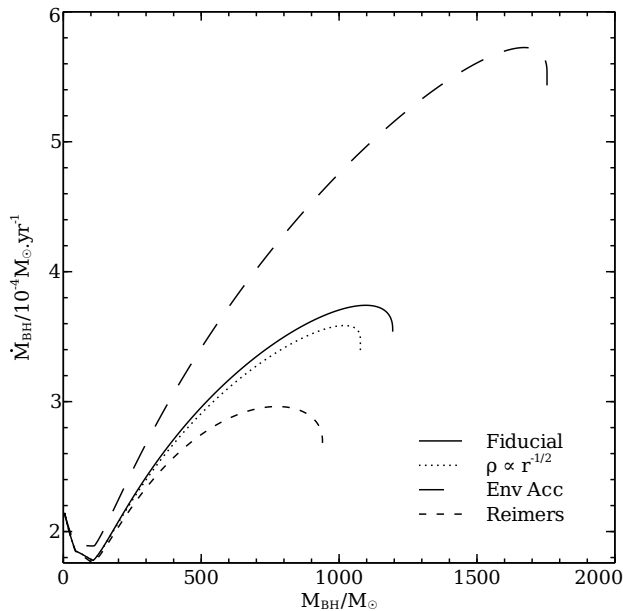}
\caption{Plot of BH accretion rate $\dMbh$ against BH mass $\Mbh$ for
the fiducial run, a run with constant accretion on to the surface of
the star (``Env Acc''), a run with a \citet{reimers+75} mass loss rate 
(``Reimers'') and a run with a shallower radial dependence of the 
interior density (``$\rho\propto r^{-1/2}$''). 
The Reimers rate leaves an envelope of about $7750\Msun$ and the BH is
proportionally smaller. The accreting envelope also leaves a 
proportionally scaled BH. A shallower radial dependence of the interior 
density leads to a smaller cavity mass and therefore a smaller interior 
density and accretion rate.}
\label{massloss}
\end{figure}

\subsection{Cavity properties}
\label{sscavprop}

The second property we adjust is the radial profile of the density 
inside the cavity. If the inward transport of angular momentum by
advection and convection is greater than the outward transport by
magnetic fields and other sources of viscosity, then the density 
profile tends towards $\rho(r)\propto r^{-\frac{1}{2}}$. Recomputing 
the cavity mass from equation (\ref{mcav}) gives
\shorteq{M_\text{cav}=\frac{8\pi}{5}\rho_0\rb^3\text{.}}
The quasi-star evolution is shown in
Fig. \ref{massloss}. The smaller interior mass leads to a less evident
density spike at all times. The decreased interior mass is subject to 
a lower mass limit for the BH. The final BH mass is $1077\Msun$.

Although the change to the cavity properties affects the numerical
results, there is no qualitative change to the quasi-star evolution.
The evolution terminates owing to the same physical
mechanism as is analysed in Section \ref{slimit}.

\subsection{Envelope mass loss and gain}
\label{ssml}

To illustrate the effect of a net accretion rate on to the surface of
the envelope, Fig. \ref{massloss} shows the evolution of the fiducial
run if the envelope is accreting at a constant rate of
$\sci{2}{-3}\Msunpyr$. Although this rate initially exceeds the
quasi-star's Eddington limit, such rapid infall is believed to occur
as long as the bars-within-bars mechanism is transporting material
towards the centre of the pregalactic cloud. Once the surface
temperature of the quasi-star decreases below about $8000\K$, the
Eddington limit becomes much larger owing to the decreasing opacity.
The accreted mass is simply added to the surface value of the mass
co-ordinate and no additions are made to any other equations. In
particular, we do not include a ram pressure at the surface. The only
qualitative change to the evolution is that it takes longer than if
the total mass had been held constant at the same final value of
$17\,540\Msun$. The final BH mass is subject to the same ratio limit
so a larger final envelope permits a larger final BH.

To investigate the effect of mass loss we use a Reimers rate
\citep{reimers+75}. It is an empirical relation to describe mass
loss in red giants. The mass-loss rate is
\shorteq{\dot{M}_\text{loss}=\sci{4}{-13}\frac{L_*R_*}{M_*}
\frac{\Msun}{\Lsun\Rsun}\text{.}}
Fig. \ref{massloss} shows the evolution for quasi-star
envelopes with this prescription. The mass loss is significant but,
again, no qualitative change in the results is seen. The limit holds
and the BH is slightly smaller, precisely in proportion with the
decrease in the mass of the envelope.

\subsection{Initial envelope mass}
\label{ssinitmass}

\begin{figure}
\includegraphics[width=84mm]{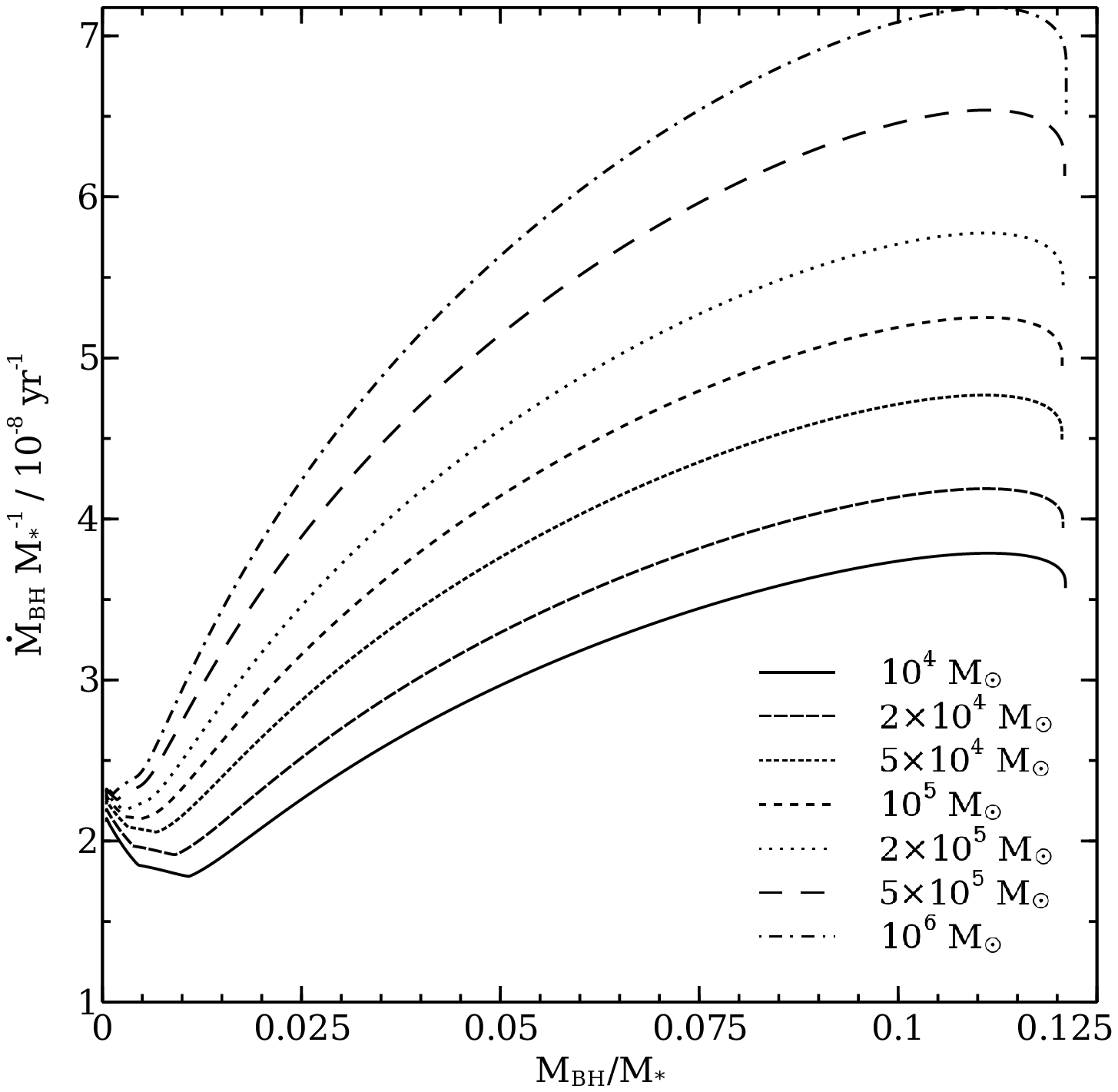}
\caption{Plot of the evolution of quasi-stars of different total
masses. The BH mass and accretion rates have been divided by
the total quasi-star mass to illustrate the consistency of the upper
mass ratio limit of $0.119$ and the slight dependence of the accretion
rate with quasi-star mass. Because larger quasi-stars permit greater
scaled accretion rates, they have shorter hydrostatic lifetimes.}
\label{minit}
\end{figure}

The structure of the envelope appears to be chiefly dependent on 
the ratio of envelope mass to 
BH mass. We constructed a set of models with initial total masses
$M_*/\Msun=10^4,\sci{2}{4},\sci{5}{4},10^5,\sci{2}{5},\sci{5}{5}$, 
and $10^6$ to confirm this and explore how the temporal evolution of 
the envelopes is affected by total mass. 

Fig. \ref{minit} shows the evolution of the various envelopes with the
BH mass and accretion rate divided by the quasi-star masses. We note 
that the fractional upper BH mass limit holds for all 
the quasi-star masses in this range. Also, the temperature profiles by 
mass of the envelopes depend almost exclusively
on the BH-envelope mass ratio. The variation in inner or surface
temperature, for a given mass ratio, with respect to the total mass
of the quasi-star is less than 0.05 per factor of ten in the total
mass.

The density and radius profiles are more strongly dependent on
the mass of the envelope. We find that, for a given mass ratio and
once the entire envelope has become convective, the following
approximate relations hold for the properties of two quasi-stars
of different masses (denoted by subscripts 1 and 2).
\shorteq{\bracfrac{R_{*,1}}{R_{*,2}}=\bracfrac{M_{*,1}}{M_{*,2}}^{0.54}\text{,}}
\shorteq{\bracfrac{\rho_{0,1}}{\rho_{0,2}}=
\bracfrac{M_{*,1}}{M_{*,2}}^{-0.66}\text{,}}
and
\shorteq{\bracfrac{\dot{M}_{\text{BH},1}}{\dot{M}_{\text{BH},2}}
=\bracfrac{M_{*,1}}{M_{*,2}}^{1.14}\text{.}}
For example, compared to a quasi-star of mass $10^4\Msun$ at the 
same BH-envelope mass ratio, a $10^5\Msun$ quasi-star will have 
an outer radius that is 
$10^{0.54}=3.47$ times greater, an interior density that is
$10^{0.66}=4.57$ times smaller and a mass accretion rate that is
$10^{1.14}=13.8$ times greater.

The final relation implies that the lifetime of a quasi-star scales as
$\tau_\text{QS}\propto M_*^{-0.14}$ so larger quasi-stars have slightly 
shorter hydrostatic lifetimes. Fig. \ref{minit} shows how the BH-envelope 
mass ratio for which the entire envelope is convective also depends on the
envelope mass. This has a small effect on the lifetime of the
quasi-stars. By fitting a straight line to the
$\log\tau_\text{QS}$--$\log M_*$ relation for the seven models here,
we find that the lifetimes scale as $\tau_\text{QS}\propto
M_*^{-0.13}$. More precisely, we find
\shorteq{\log_{10}\tau_\text{QS}=-0.126\log_{10}M_*+7.12}
Note that $M_*$ here denotes the initial mass of the quasi-star. In
all other relations $M_*$ slowly decreases during the quasi-star's evolution
owing to the mass-energy lost as radiation.

\subsection{Comparison with Begelman et al. (2008)}
\label{sbra}

\begin{figure}
\includegraphics[width=84mm]{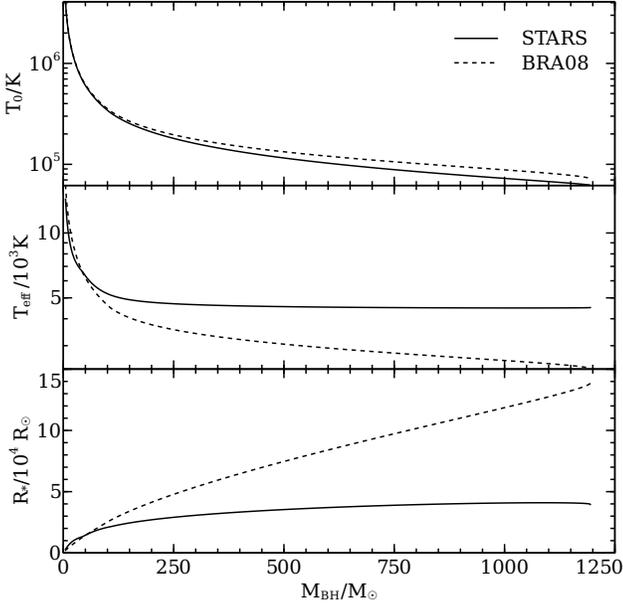}
\caption{Comparison of analytic estimates of BRA08 (dashed lines)
against results for our fiducial run (solid lines) for interior
temperature (top), surface temperature (middle) and envelope radius
(bottom). BRA08's estimate of the interior temperature is accurate
but those for the photospheric temperature and envelope radius become
increasingly inaccurate as the BH grows.}
\label{brac}
\end{figure}

\begin{figure}
\includegraphics[width=84mm]{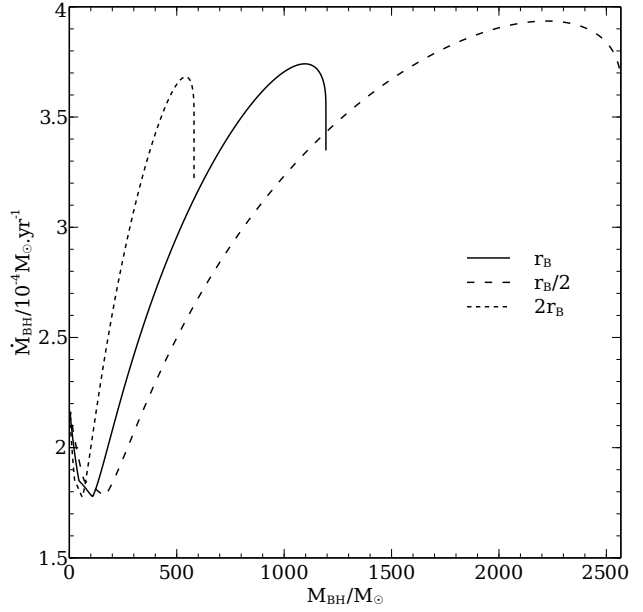}
\caption{Plot of BH accretion rate $\dMbh$ against BH mass $\Mbh$ for
different choices of the inner boundary radius. The various values of
$r_0$ lead to qualitatively similar results but the quantitative
evolution is strongly affected.}
\label{rsense}
\end{figure}

BRA08 have estimated some envelope properties presuming that the
envelope is described by an $n=3$ polytrope. Their estimates are made
in terms of an overall accretion efficiency parameter\footnote{This
should not be confused with $\alpha$ defined in \mbox{Section
\ref{sslp}}.} $\alpha_\text{BRA}$ which is determined by
numerical factors in the accretion rate including the radiative
efficiency, convective efficiency and adiabatic index. Because our
adiabatic index is not fixed, we have calibrated $\alpha_\text{BRA}$
using the BH luminosity (equation 3 of BRA08). To compare the fiducial
run, we selected $\alpha_\text{BRA}\approx0.257$ which makes their
analytical BH luminosity accurate to within 0.2 per cent over the
entire evolution.

We compare the following estimates for the inner temperature,
photospheric temperature and envelope radius (equations 7, 11 and 10
of BRA08, respectively).
\shorteq{T_0=\sci{1.4}{4}\left(\frac{L}{L_\text{Edd}}\right)^\frac{2}{5}
\left(\alpha_\text{BRA} \frac{\Mbh^2}{\mathrm{M}_\odot^2}\right)^{-\frac{2}{5}}
\left(\frac{M_*}{\Msun}\right)^\frac{7}{10}\K\text{,}}
\shorteq{T_\text{eff}=\sci{1.0}{3}\left(\frac{L}{L_\text{Edd}}\right)^\frac{9}{20}
\left(\alpha_\text{BRA} \frac{\Mbh^2}{\mathrm{M}_\odot^2}\right)^{-\frac{1}{5}}
\left(\frac{M_*}{\Msun}\right)^\frac{7}{20}\K}
and
\shorteq{R_*=\sci{4.3}{14}\left(\frac{L}{L_\text{Edd}}\right)^{-\frac{2}{5}}
\left(\alpha_\text{BRA} \frac{\Mbh^2}{\mathrm{M}_\odot^2}\right)^\frac{2}{5}
\left(\frac{M_*}{\Msun}\right)^{\frac{1}{5}}\cm\text{.}}
Here, $L_\text{Edd}=4\pi GcM/\kappa$
is the Eddington luminosity. BRA08 computed this using the
opacity at the boundary of the convective zone but such estimates 
differ by a factor of the order of $\kappa/\kappa_\text{es}$ when
compared with our results. Our comparison is made using the
Eddington limit with opacity $\kappa_\text{es}=0.34\cmsq\pg$.

In Fig. \ref{brac}, we plot the three estimates against the results
from our fiducial run. The estimate for the interior temperature is
accurate to within 20 per cent. The deviation grows as the
approximation of the envelope to an $n=3$ polytrope becomes
increasingly less accurate.

The estimate for the photospheric temperature is highly inaccurate. At
the end of the run the estimated photospheric temperature is about
$2400\K$ compared to the model result of about $4500\K$. Because the BH
luminosity estimate is very accurate, it then follows that the
envelope radius is inaccurate. The surface luminosity must be
related by $L_*=\pi acR_*^2T_\text{eff}^4$. This is confirmed in the
bottom panel of Fig. \ref{brac}.

BRA08 argue that quasi-star evolution terminates owing to the
opacity at the edge of the convection zone increasing. The increased
opacity causes the envelope to expand and the opacity increases further. 
The envelope then expands further and the process is claimed 
to run away. BRA08 refer to this process as the \emph{opacity crisis}. Our
results do not terminate for this reason. Similar behaviour does occur at the
beginning of the evolution while the photospheric temperature is
greater than $10^4\K$ but it does not disperse the quasi-star. For
most of a quasi-star's evolution, the opacity at the convective
boundary is already beyond the H-ionization peak and is decreasing as
the BH grows.

\section{Sensitivity to the inner boundary}
\label{sbound}

\begin{figure}
\includegraphics[width=84mm]{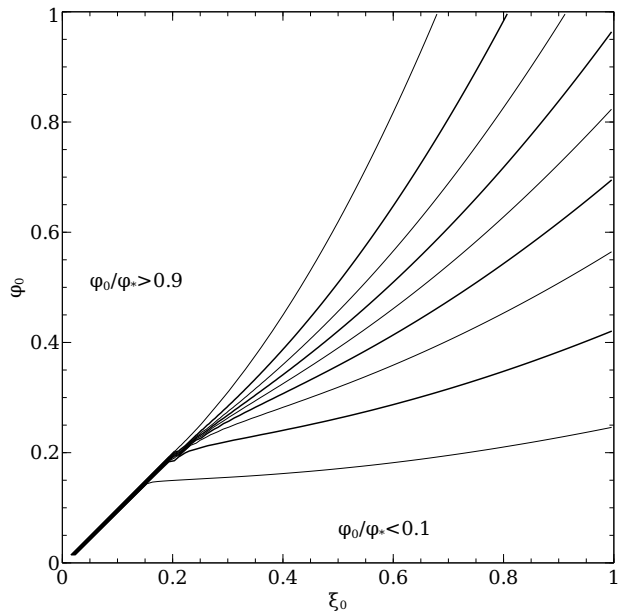}
\caption{Contours of $\phi_0/\phi_*$ for the models calculated for
each pair of initial conditions $(\xi_0,\phi_0)$ in the plane. There 
is a clear boundary along $\phi_0=\xi_0$ near the origin. Solutions 
above this line must have negligibly small envelopes by mass.}
\label{pmxf}
\end{figure}

Besides the physical parameters of the models, we have also considered
the effect of changing the inner boundary radius. Fig. \ref{rsense}
shows the evolution of quasi-stars where the inner radius was changed
to half and twice the Bondi radius. The final BH mass is strongly
affected although the evolution is qualitatively similar. The results
of our polytropic analysis in Section \ref{slimit} are also affected
in a consistent manner. Although reasonable, our choice of inner
radius is somewhat arbitrary and critical in deciding the quantitative
evolution of the quasi-star. A self-consistent model to determine the
appropriate value of $r_0$ is thus highly desirable but none is yet
available. We continue to seek a suitable resolution to this issue.

\begin{figure}
\includegraphics[width=84mm]{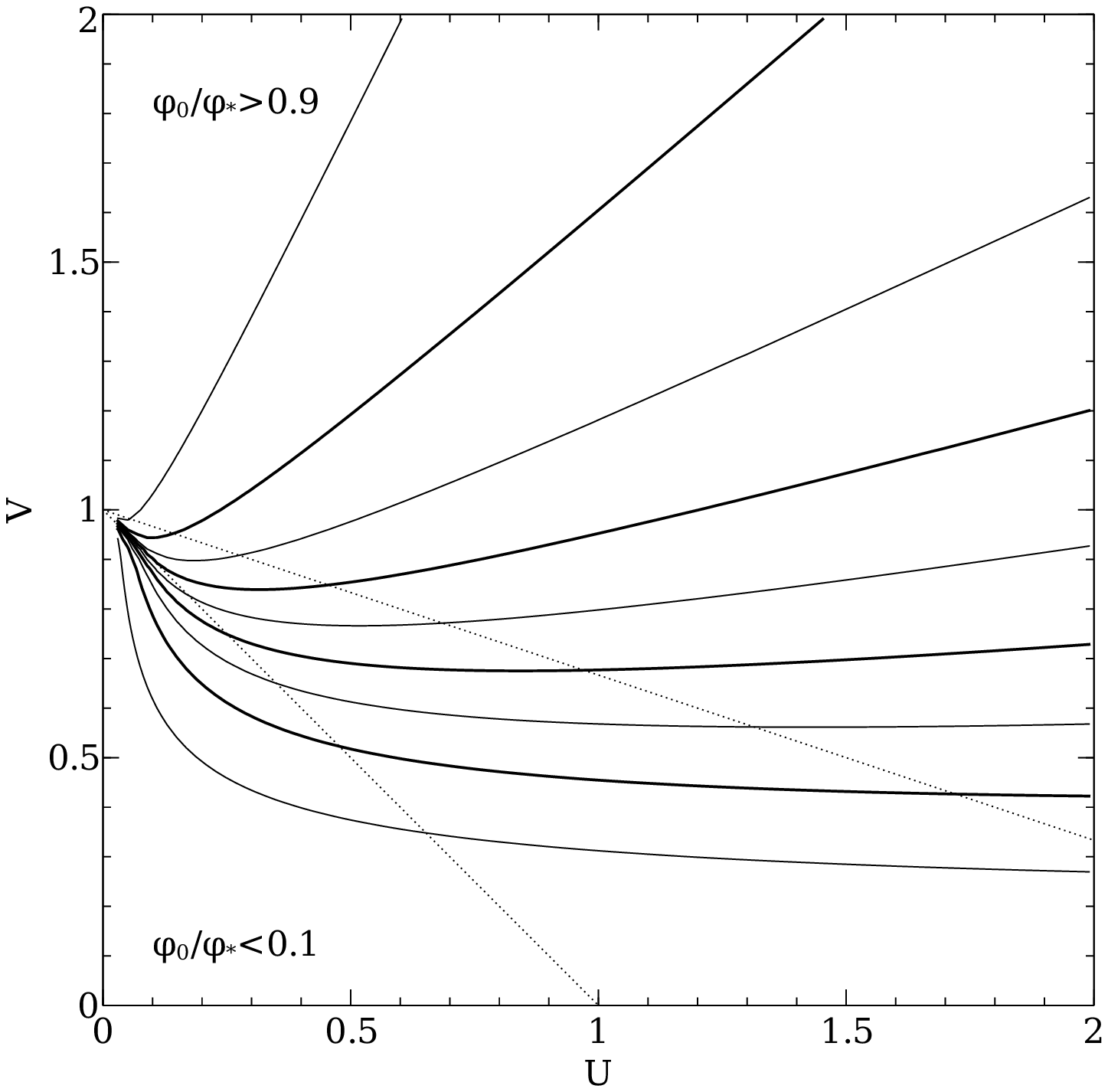}
\caption{Contours of $\phi_0/\phi_*$ for the models calculated for
each pair of initial conditions $(U_0,V_0)$ in the plane. The boundary
in Fig. \ref{pmxf} corresponds to the critical point (0,1).}
\label{pmuv}
\end{figure}

The sensitivity of the models to the inner boundary precipitated
further interesting results. We note that the final BH mass is scaled
by approximately the same factor as the inner radius. That is, if
$r_0$ is doubled, the BH reaches about half the final mass that it
reached before. If the material properties at the inner boundary were
largely unchanged, this would imply that the inner radius at the
termination of the runs is approximately the same. The output from the
models indicates that $r_0/\Rsun$ in the final reliable model of each
run in Fig. \ref{rsense} is approximately $\sci{2.05}{4}$,
$\sci{1.94}{4}$ and $\sci{1.79}{4}$ for $r_0=2\rb$, $\rb$ and
$\rb/2$. This indicates that the limit determined for the polytropic
models in Section \ref{slimit} may be related more strongly to the
ratio of inner to outer radius than the ratio of interior to total
mass.

It may appear from Fig. \ref{matlab} that the limiting radius ratios
for the STARS output and the polytropic models are inconsistent. The
discrepancy is due to the ionization zones inside the envelope. It is
also difficult to identify the appropriate values because the inner
and outer radii are changing most rapidly at the end of the
evolution. Using the last models in each run of Section
\ref{ssinitmass} that have converged in thermal equilibrium, the limit
is approximately $r_0/R_*\approx0.460$.

We additionally found that we could not construct model envelopes for
$r_0\la0.3\rb$ and find that this is reflected in the polytropic
models. To construct Fig. \ref{pmxf} we have calculated the ratio
$\phi_0/\phi_*$ for each $(\xi_0,\phi_0)$ pair in the plane and
plotted contours from $0$ to $1$ in steps of $0.1$. Near the origin,
there is a clear boundary along the line $\phi_0=\xi_0$. The boundary
softens for larger values of $\phi_0$ and $\xi_0$. For small $\xi_0$,
models above the boundary necessarily have $\phi_0/\phi_*\approx1$. In
other words, any model with a sufficiently small inner boundary must
have an appropriately small envelope.

The limit becomes clearer when plotted in the $U$--$V$ plane discussed
in Section \ref{ssuv}. We have done so in Fig. \ref{pmuv}. The
contours are divided across the point (0,1). This is a critical point
in the $U$--$V$ plane for all polytropic indices. It is neutrally
stable in the direction (1,-1) and unstable along the $V$-axis
i.e. (0,1).

The present analysis thus provides several important results. First,
the choice of inner radius is crucial to determining the evolution of
the envelopes. The Bondi radius is a reasonable choice but there is no
evidence to confirm that it is the correct one. Secondly, the limit
determined in Section \ref{slimit} seems more strongly related to the
radius ratio than the mass ratio. This was not obvious at first
because there was no apparent reason to modify the inner radius.
Thirdly, there is a limit to the smallest value that can be chosen for
the inner radius.

\section{Conclusion}
\label{sconc}

We have modified the Cambridge \stars{} stellar evolution code to
model the evolution of quasi-star envelopes. Our first new result is
the existence of a robust upper limit on the ratio of inner BH mass to
the total mass, equal to about $0.119$, of the system in hydrostatic
equilibrium. The limit is reflected in solutions of the Lane-Emden
equation, modified for the presence of a point mass interior to some
specific boundaries. After considering variation of the inner radius,
this limit is possibly better interpreted as a limit to the ratio of
inner radius to envelope radius. The value of the limit from the STARS
output is about $0.46$.

All the evolutionary runs here terminate once the limit is reached. It
is difficult to say what happens to the BH and envelope after the
hydrostatic evolution ends. Some of the material within the Bondi
radius has begun accelerating towards the BH so we expect that it can
be captured by the BH. The remaining material may be accreted or
expelled, depending on the liberation of energy from the material that
does fall inwards. After the BH has evolved through the quasi-star
phase, it is probably limited to accreting at less than the Eddington
limit for the BH.

The models presented are crucially sensitive to the choice of inner
boundary radius and the results should be treated with due
caution. While the Bondi radius used here is reasonable, we continue
to seek a less arbitrary set of boundary conditions.

In light of these results, it appears that quasi-stars produce BHs
that are on the order of at least $0.1$ of the mass of the quasi-star
and around $0.5$ if all the material within the inner radius is
accreted. For conservative parameters, this growth occurs within a few
million years after the BH initially forms. Realistic variations in
the parameters (e.g. larger initial mass, lower radiative efficiency)
lead to shorter lifetimes. Such BHs could easily reach masses
exceeding $10^9\Msun$ early enough in the Universe to power
high-redshift quasars.

\section*{Acknowledgements}

The authors would like to thank Peter Goldreich for valuable comments.
We are grateful to Mitch Begelman, particularly for comments 
that led to the investigation in Section \ref{sbound}.
CAT thanks Churchill College for a Fellowship.
\balance
\footnotesize{

}
\bsp
\end{document}